# High-symmetry ill-fitting subunits in 3D form aggregates of all dimensions

Elena N. Govorun[1] and Martin Lenz[1,2]
[1] *LPTMS, UMR 8626 CNRS, University Paris-Saclay, Orsay, 91400 France*
[2] *PMMH, UMR 7636 CNRS, ESPCI Paris-PSL, Sorbonne Université, Université de Paris, Paris, 75005 France*

Proteins can combine into functional elements in living cells or self-assemble into unwanted structures in a number of diseases. The resulting aggregates often display filamentous morphologies across a large range of protein shapes and molecular interactions. This has led to the suggestion that filament formation could be a generic outcome of the aggregation of geometrically complex, ill-fitting objects, although such a mechanism has not been demonstrated in three dimensions. To address this problem, we theoretically study the self-assembly of three-dimensional identical, ill-fitting deformable subunits mimicking globular proteins in solution. In our model, self-assembling subunits incur deformations that accumulate as the aggregate size increases and can eventually hamper further assembly. We analytically predict the ground state morphologies of the resulting aggregates as a function of the subunit adhesivity and elasticity by mapping their mechanics onto those of two incompatible, interconnected networks. We find that zero-dimensional clusters, three-dimensional bulks as well as symmetry-broken one-dimensional filaments and two-dimensional layers can all form depending on assembly parameters. Poorly compressible, moderately adhesive subunits favor filaments. These findings hint at a generic pathway to control self-assembly in three dimensions and suggests that such mechanisms could be investigated in more realistic protein models.

## 1. Introduction

The self-assembly of subunits into aggregates is common in living and nonliving systems. It drives the formation of viral capsids [1-3], as well as of filamentous protein aggregates observed in many diseases [4-10]. Understanding the underlying mechanisms could have biomedical implications, and both experimental and theoretical investigations into them are routinely conducted using artificial subunits with simplified geometries [11,12].

Self-assembly can also be harnessed for nanotechnological applications. The technology of DNA origami thus allows the design and manufacture of three-dimensional DNA nanoparticles with a very wide range of shapes. These nanoparticles can then be further self-assembled into higher-order structures such as truncated octahedra, artificial capsids, and DNA-linked colloidal clusters [13-23]. Designing such higher-order structures however remains a challenge. Indeed, although any aggregate morphology can in principle be assembled out of unique subunits with specific interactions, a design strategy known as fully programmable assembly [18-21]. Such an approach is however extremely costly in practice, which motivates a search for physical effects that help control the morphology of aggregates comprised of identical subunits [22-26].

The morphology of the aggregates formed upon self-assembly depends on the shape of the assembling subunits. Adhesive isotropic subunits often form homogeneous, arbitrarily large macroscopic aggregates. Conversely, anisotropic subunits such as rod-like viruses or oligopeptides, may self-assemble into tapes, twisted ribbons or helices whose width saturate to a finite value [27,28]. More generally, a misfit in the subunit shapes can cause self-assembly to stop after the aggregates have reached a certain size as accumulating internal stresses penalize further growth [12,29-36].

Such stress accumulation can cause two-dimensional subunits with orientation-dependent, specific interactions to forgo isotropic growth and form tape-like aggregates [30-32]. Similarly, stress accumulation in polygonal deformable subunits in two dimensions with nonspecific interactions may preclude the formation of a bulk phase and lead to the formation of disk- or stripe-like aggregates [37,38]. Such deformability may have an analog in proteins, where conformational changes are involved in molecular transport and in protein folding [39,40], and are additionally revealed in mutants where the stability of a native protein state is perturbed [41]. Protein deformations are additionally involved in the regulation of viral self-assembly [1,2] and in allostery [42,43]. Although a combination of attractive and repulsive interactions in tetrahedral subunits has been shown to lead to the formation of helical ribbons in three dimensions [36], the relative stability of such structures compared to finite-size aggregates of other dimensions as well as their possible emergence in systems of deformable subunits have not been systematically investigated.

Here we develop a theoretical formalism based on the elasticity of simple ill-fitting subunits to determine which aggregate morphology among zero-dimensional clusters, one-dimensional filaments, two-dimensional layers or three-dimensional bulks is more energetically favorable in self-assembly in three-dimensional space. In Sec. 2, we present our continuum elastic model based on two coupled elastic networks, and illustrate the type of subunit design that can form its microscopic foundation. In Sec. 3 we compute the distribution of strains, stresses, and energies for the morphologies listed above. In Sec. 4 we compare the corresponding energies to derive a morphological diagram of the 3D aggregation process depending on networks' Poisson's ratio and the subunits' adhesive strength. We find that subunits with low Poisson's ratios undergo a discontinuous transition from zero-dimensional clusters to bulk, while for high enough Poisson's ratio the dimensionality of the most favorable aggregate increases monotonically as the subunit adhesivity is increased. While the phase boundaries of this diagram cannot be expressed in a closed form with usual functions, in Sec. 5 we consider the analytically simpler limit of almost-incompressible (high Poisson's ratio) subunits and derive an explicit form for these boundaries. Finally we consider a generalization of our model to dimensions higher than three in Sec. 6, and derive the phase boundaries in the almost-incompressible limit in this case also, with conclusions similar to those of the three-dimensional case. We discuss our results in Sec. 7.

## 2. Model

We consider a set of ill-fitting subunits that self-assemble in three-dimensional space. Following the example of Ref. 38, we describe the surface of the subunits by keeping track of a set of points located on their surface, *e.g.*, protein functional groups. A schematic example of such a morphology is displayed in Fig. 1. First consider the case where well-adjusted subunits fit perfectly together to form a three-dimensional crystal-like aggregate without any subunit deformation. In that case each set of points lies on the same crystalline lattice with the same lattice spacing. By contrast, here we consider a minimal model of subunits that cannot fit together without deformation. We thus consider subunits with two sets of points. We assume that each set still prefers sitting on its crystalline lattice, only now the two lattices have different lattice spacings. We denote by $2\epsilon$ the relative difference between these spacings.

We consider an aggregate formed by a large enough number of subunits to be described by a continuum formalism. In this formalism, we keep track of the displacements of the two sets of points, which we refer to as the "yellow" and "red" points through two displacement fields $\mathbf{u}^{(\mathrm{y})}(\mathbf{r})$ and $\mathbf{u}^{(\mathrm{r})}(\mathbf{r})$, where $\mathbf{r}$ denotes the position in three-dimensional space. The elastic energy of the system is a functional of these two fields. We define the reference state of the continuum material as the elastic ground state of an infinite aggregate, meaning that the elastic energy of a bulk aggregate is minimal when both fields uniformly vanish.

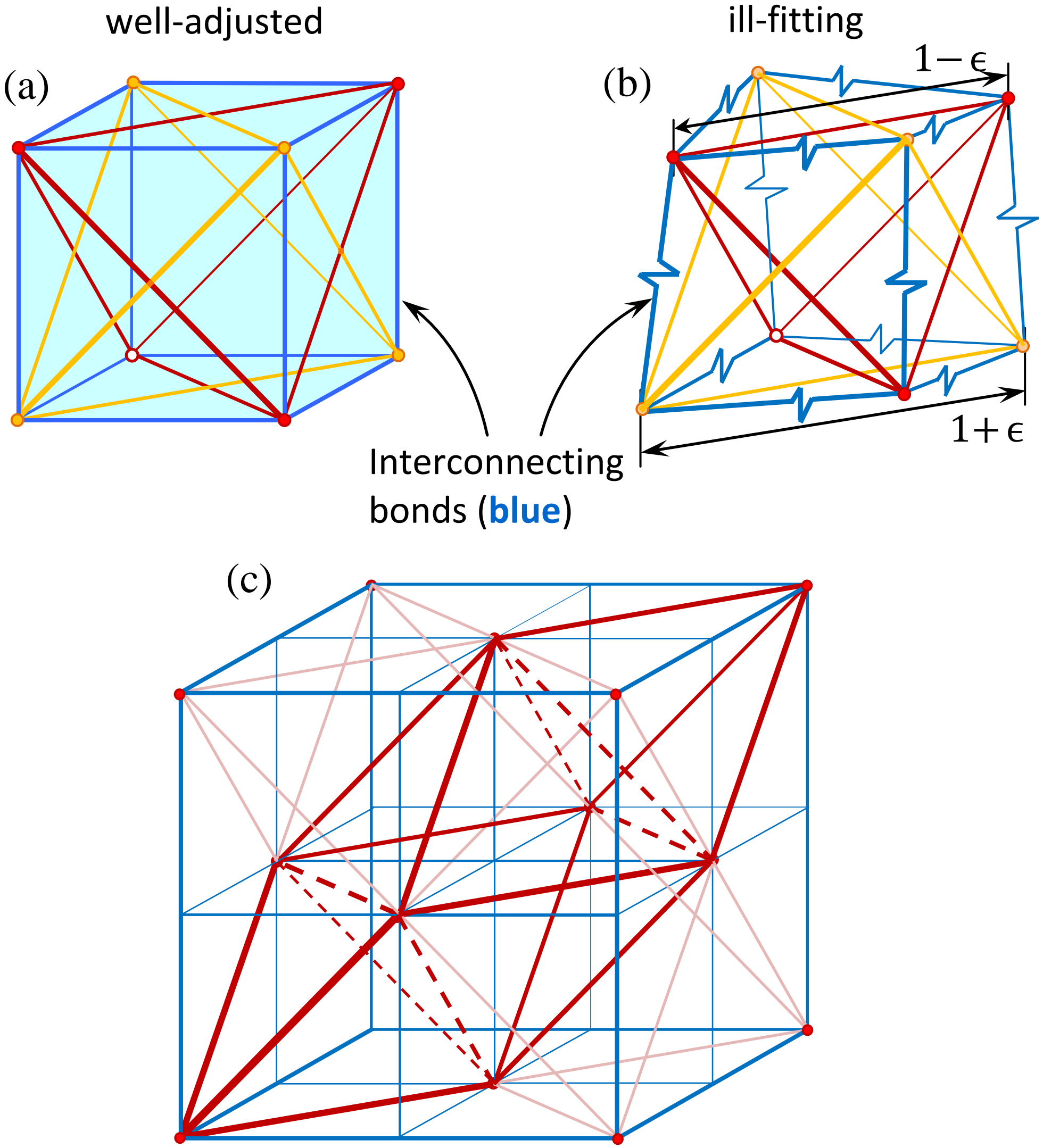

Fig. 1. Illustration of the subunit geometry that gives rise to the continuum model studied here. Note that our continuum models is built on the basis of symmetry considerations and is thus not restricted to this specific design. (a) Case of well-adjusted subunits. Here each subunit is a cube with its vertices alternately colored yellow and red. The yellow set of points and the red set of points each form a regular tetrahedron whose edges act as springs. Neighboring points of different colors are connected by blue springs. (b) In the case of ill-fitting subunits, the yellow and red tetrahedra have slightly different sizes, which deforms isolated subunits into cuboids. (c) We consider an aggregation rule such that yellow points stick together, and similarly for red ones. Forming an infinite aggregate thus requires deforming each subunit into a perfect cube, then packing these cubes together. For clarity here we display only the red and blue springs of the resulting lattice; the red network forms a face-centered cubic lattice.

We construct our model as the most symmetric model compatible with the phenomenology discussed above. Our free energy density is thus defined as the most general isotropic, translationally invariant quadratic elastic functional of the two displacement fields that is moreover invariant upon simultaneously exchanging the red and yellow lattice and switching the sign of $\epsilon$. While this model forms the continuum limit of the design of Fig. 1, it is much more general than this specific example. Under this set of assumptions, our free energy density can be written as the sum of three terms. The first two can be viewed as the elastic energy densities $e_{\mathrm{y}}(\mathbf{r})$ and $e_{\mathrm{r}}(\mathbf{r})$ of the two individual networks:

$$e_{\mathrm{y}} = \frac{\lambda}{2}\left(\partial_\alpha u_\alpha^{(\mathrm{y})} - 3\epsilon\right)^2 + \mu\left(\frac{\partial_\alpha u_\beta^{(\mathrm{y})} + \partial_\beta u_\alpha^{(\mathrm{y})}}{2} - \epsilon\delta_{\alpha\beta}\right)^2, \tag{1}$$

$$e_{\mathrm{r}} = \frac{\lambda}{2}\left(\partial_\alpha u_\alpha^{(\mathrm{r})} + 3\epsilon\right)^2 + \mu\left(\frac{\partial_\alpha u_\beta^{(\mathrm{r})} + \partial_\beta u_\alpha^{(\mathrm{r})}}{2} + \epsilon\delta_{\alpha\beta}\right)^2, \tag{2}$$

where $\lambda$ and $\mu$ can be viewed as the Lamé coefficients of the individual networks and the summation over repeated indices is implied. The indices $\alpha$ and $\beta$ denote the spatial coordinates, and $\delta_{\alpha\beta}$ is the Kronecker delta. The only dimensionless parameter characterizing the elasticity of the networks is their Poisson's ratio $\nu = \lambda/(2\lambda + 2\mu)$. A larger Poisson's ratio indicates a less compressible network, up to a maximum value of $\nu = 1/2$ which corresponds to a fully incompressible medium. These two single-network terms are minimal when the yellow and red networks strain isotropically by an amount $\epsilon$ and $-\epsilon$ respectively, implying that the colored networks incur residual stresses in the reference bulk state.

The third term can be thought of as the energy per unit volume of the blue interconnecting bonds of Fig. 1, namely

$$e_{\mathrm{inter}} = \frac{\kappa_c}{2}\left[\mathbf{u}^{(\mathrm{y})}(\mathbf{r}) - \mathbf{u}^{(\mathrm{r})}(\mathbf{r})\right]^2, \tag{3}$$

where the constant $\kappa_c$ is the stiffness of the interconnection between the two networks. These interconnections dominate large aggregates, and thus forces the alignment of the red and yellow network in the reference state. Indeed, any mismatch between the displacement gradient tensors $\partial_\alpha u_\beta^{(\mathrm{y/r})}$ of the red and yellow strain tensors would lead to an energy density $e_{\mathrm{inter}}$ that scales as the square of the lateral aggregate size. This is prohibitive in infinite aggregates.

The elastic cost of aggregating the subunits can be partially compensated by the subunits' adhesion energy. We can express it as a surface tension-like surface energy that is proportional to the area $S$ of the outer surface of the aggregate, which in contact with the surrounding solvent

$$E_{\mathrm{surf}} = \gamma S, \tag{4}$$

where $\gamma$ denotes the surface tension.

The energy of an aggregate is a sum of all the terms discussed above, namely

$$E = \int dV\left(e_{\mathrm{y}} + e_{\mathrm{r}} + e_{\mathrm{inter}}\right) + \gamma S. \tag{5}$$

Here the integration volume $V$ and the surface area $S$ are evaluated in the reference state where the displacement fields of the two networks vanish.

## 3. Study of the putative aggregate morphologies

Our aggregation model allows us to consider the energetic costs and benefits of forming different aggregate morphologies. In Sec. 3.1 we first consider the reference case of a bulk aggregate, then discuss the mechanical equilibrium equation in arbitrary geometries. We then study the various self-limited morphologies considered here by increasing order of computational complication, namely slabs (Sec. 3.2), filaments (Sec. 3.3) and spheres (Sec. 3.4). We furthermore consider the case of hollow spherical shells (vesicles) in the *Supplementary Information*, but find that they are never more stable than the cases listed above.

### 3.1 Bulk state and mechanical equilibrium

In the bulk state, the yellow and red lattices are perfectly aligned, implying a vanishing interconnection energy. The surface of the aggregate is moreover negligible, implying that the energy per unit volume is equal to the sum of the energies of the two elastic networks $e_{\mathrm{y}}$ and $e_{\mathrm{r}}$ [Eqs. (1) and (2)] at $\mathbf{u}^{(\mathrm{y})} = \mathbf{u}^{(\mathrm{r})} = 0$, namely

$$e_{\mathrm{bulk}} = 3\epsilon^2(3\lambda + 2\mu). \tag{6}$$

In finite aggregates, the subunits can partially relieve these elastic energies by stretching the interconnecting springs in the vicinity of the aggregate surface. Minimizing the free energy functional relative to the displacement fields of the yellow and red network respectively gives the force balance equation for each separate network

$$\partial_\beta \sigma_{\alpha\beta}^{(\mathrm{y})} = f_\alpha^{(\mathrm{y})} = \kappa_c\left(u_\alpha^{(\mathrm{y})} - u_\alpha^{(\mathrm{r})}\right) \tag{7}$$

$$\partial_\beta \sigma_{\alpha\beta}^{(\mathrm{r})} = f_\alpha^{(\mathrm{r})} = -\kappa_c\left(u_\alpha^{(\mathrm{y})} - u_\alpha^{(\mathrm{r})}\right),$$

where the interconnecting blue bonds appear as external forces $f_\alpha^{(\mathrm{y})} = \partial e_{\mathrm{inter}}/\partial u_\alpha^{(\mathrm{y})}$, $f_\alpha^{(\mathrm{r})} = \partial e_{\mathrm{inter}}/\partial u_\alpha^{(\mathrm{r})}$. The elements of the stress tensors in Eq. (7) are given by the networks' constitutive relations

$$\sigma_{\alpha\beta}^{(\mathrm{y})} = \frac{\partial e_{\mathrm{y}}}{\partial(\partial_\alpha u_\beta^{(\mathrm{y})})} = \lambda\left(\partial_\gamma u_\gamma^{(\mathrm{y})} - 3\epsilon\right)\delta_{\alpha\beta} + \mu\left(\partial_\alpha u_\beta^{(\mathrm{y})} + \partial_\beta u_\alpha^{(\mathrm{y})} - 2\epsilon\delta_{\alpha\beta}\right) \tag{8}$$

$$\sigma_{\alpha\beta}^{(\mathrm{r})} = \frac{\partial e_{\mathrm{r}}}{\partial(\partial_\alpha u_\beta^{(\mathrm{r})})} = \lambda\left(\partial_\gamma u_\gamma^{(\mathrm{r})} + 3\epsilon\right)\delta_{\alpha\beta} + \mu\left(\partial_\alpha u_\beta^{(\mathrm{r})} + \partial_\beta u_\alpha^{(\mathrm{r})} + 2\epsilon\delta_{\alpha\beta}\right).$$

These internal force balance equations are supplemented by a zero-stress boundary condition at the aggregate's free surface

$$n_\alpha \sigma_{\alpha\beta}^{(\mathrm{y})} = n_\alpha \sigma_{\alpha\beta}^{(\mathrm{r})} = 0, \tag{9}$$

where $\mathbf{n}(\mathbf{r})$ is the normal vector to the surface. In the following we use Eqs. (7) and (9) to derive the displacements and stresses corresponding to each morphology, then compute the overall aggregate energy $E$ using Eq. (5).

### 3.2 Flat slabs

We consider a slab of fixed thickness $W$ and first compute the corresponding aggregate energy $E$ as a function of $W$. We then derive the optimal value of $W$ by minimizing the system's total energy. We take our slab as orthogonal to the $x$-axis with its midplane at $x = 0$. The displacements of the yellow and red networks vanish in the $yz$ plane. Indeed, any other choice of constant strain would result in a diverging elastic energy. We define the displacement difference $\delta(x) = u^{(\mathrm{y})}(x) - u^{(\mathrm{r})}(x)$, which allows us to write the force balance of Eq. (7) as

$$(\lambda + 2\mu)\frac{\partial^2 \delta(x)}{\partial x^2} = 2\kappa_c \delta(x). \tag{10}$$

The problem's symmetry with respect to the $x = 0$ plane imposes $u^{(\mathrm{y})/(\mathrm{r})}(x) = -u^{(\mathrm{y})/(\mathrm{r})}(-x)$, which implies that $u^{(\mathrm{y})/(\mathrm{r})}(0) = 0$, and $\delta(0) = 0$. As a result, the solution of Eq. (10) has the form

$$\delta(x) = A \cdot \sinh\frac{x}{l} \qquad \text{with} \qquad l = \sqrt{\frac{\lambda + 2\mu}{2\kappa_c}}, \tag{11}$$

where $A$ is a constant to be determined. As a result of symmetry the yellow and red displacements differ only by their signs: $u^{(\mathrm{y})}(x) = -u^{(\mathrm{r})}(x) = \delta(x)/2$.

The bulk Eqs. (10-11) are identical to the equations previously derived for a one-dimensional strip-like aggregate of finite width in two dimensions [38]. They however satisfy a different boundary condition, namely the cancellation of the three-dimensional longitudinal stress component $\sigma_{xx}^{(\mathrm{y})/(\mathrm{r})} = \frac{\partial e_0}{\partial\left(\partial_x u_x^{(\mathrm{y})/(\mathrm{r})}\right)} = 0$ at $x = \pm W/2$, which yields

$$A = \frac{2l\epsilon(3\lambda + 2\mu)}{(\lambda + 2\mu)\cosh W/(2l)}. \tag{12}$$

While the interconnections described by Eq. (3) control large aggregates, small-scale deformations are dominated by the elastic energies of Eqs. (1-2), which are higher order in gradient. The length $l$ introduced in Eq. (11) describes the characteristic length scale over which these contributions are of the same order. The continuum representation of Eqs. (1-5) is strictly valid only when this length scale is much larger than the size of a subunit. This length can also be expressed as $l = \sqrt{M/(2\kappa_c)}$, where $M = \lambda + 2\mu$ is the P-wave, or longitudinal, modulus.

The $x$-dependence of the displacement difference $\delta(x)$ and stress tensor element $\sigma_{xx}^{(\mathrm{r})}$ are presented in Fig. 2. For aggregates whose size is large with respect to the characteristic length $l$, the displacements $\mathbf{u}^{(\mathrm{y})}, \mathbf{u}^{(\mathrm{r})}$ are small in most of the slab's volume (as in a large bulk aggregate), but become substantial in two exponentially decaying boundary layers located at the aggregate surface. The intra-network stresses are maximum at the center of the slab and equal to zero at the surface as required by the boundary condition. The overall magnitude of the displacements and increase significantly when Poisson's ratio is increased while holding the shear modulus $\mu$ fixed. For aggregate sizes smaller than or comparable to $l$, the displacement difference increases almost linearly with $x$ (Fig. 2a, solid lines).

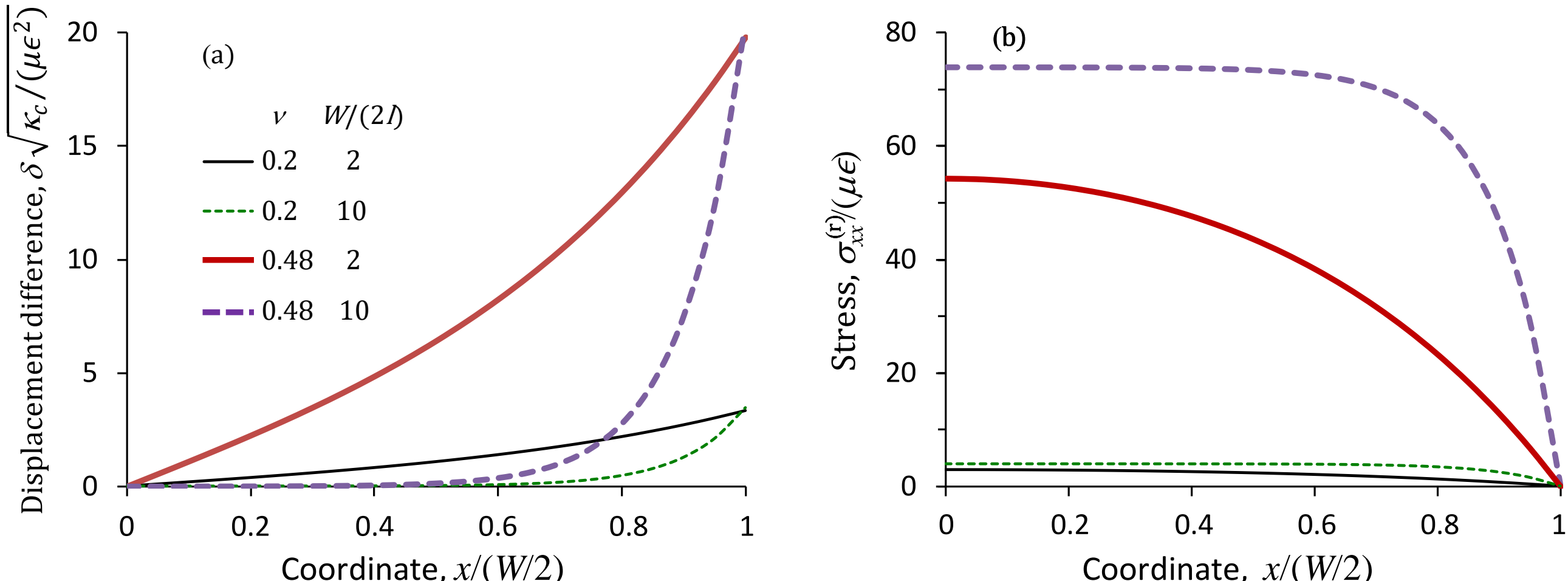

Fig. 2. Displacement and stress in two-dimensional slabs at mechanical equilibrium. (a) Difference $\delta(x)$ in the displacements of the networks and (b) stress tensor element $\sigma_{xx}^{(r)}$ in a half of a slab ($0 \le x \le W/2$). Note that the characteristic length $l$ increases with increasing Poisson's ratio: $l_{\nu=0.48} \approx 3l_{\nu=0.2}$.

The surface energy of a slab of area $S_2$ is:

$$E_{\mathrm{surf2}} = 2\gamma S_2 = 2\gamma\frac{V}{W}, \tag{13}$$

where the index "2" here and in the following is associated with the slab, *i.e.*, an aggregate infinite in two directions. By substituting the solution Eq. (11-12) in the expressions for the aggregate energy [Eqs. (1-5)], we find an energy per unit volume of the slab $e_2(W)$ of the form

$$\frac{e_2(W)}{e_{\mathrm{bulk}}} = 1 - \frac{2}{3}\frac{1+\nu}{1-\nu}\left(\frac{\tanh\frac{W}{2l}}{W/l} - \frac{\Gamma}{W/l}\right), \tag{14}$$

where we define the dimensionless surface tension as

$$\Gamma = \frac{3\gamma(1-\nu)}{e_{\mathrm{bulk}}l(1+\nu)}. \tag{15}$$

This parameter is the ratio of the typical surface and bulk energies of a slab of thickness $l$.

To find the optimum value of $W$, we reason that aggregation proceeds through the rearrangement of a fixed number of subunits in solution. Minimizing the total energy of the system is therefore equivalent to minimizing the energy per subunit, which is again equivalent to minimizing the energy per unit of reference volume $e_2(W) = E/V$ with respect to $W$. For $\Gamma \geq 1$, the energy Eq. (14) decreases monotonically with increasing thickness $W$, implying that aggregating into an infinite bulk is more energetically favorable than forming a slab. For $\Gamma \leq 1$, the energy per volume has an absolute minimum and the equilibrium slab thickness $W^*$ can be calculated from the condition $\partial e_2(W)/\partial W = 0$. This equation is equivalent to the following condition on the dimensionless width $s = W^*/l$

$$\frac{s}{2}\frac{1}{\cosh^2 s/2} - (\tanh s/2 - \Gamma) = 0, \tag{16}$$

which is transcendental and therefore has no simple solution in terms of usual functions. For small values of $\Gamma$, Eq. (16) yields an asymptotic equilibrium thickness $W^*/l \sim \Gamma^{1/3}$. As $\Gamma$ increases, the slab thickness increases and it tends to infinity as $-\log(1-\Gamma)$ for $\Gamma \to 1^-$. Inserting the value of the optimal thickness into Eq. (14) yields the optimal slab energy per unit volume.

### 3.3 Filaments

We model a filament as an infinite cylinder of radius $R$ with the $z$ axis as its centerline. The network displacements $u^{(\mathrm{y})}(r)$, $u^{(\mathrm{r})}(r)$ depend only on the distance $r$ from that axis. The force balance Eq. (7) yields

$$\left(\frac{\partial^2 \delta(r)}{\partial r^2} + \frac{1}{r}\frac{\partial \delta(r)}{\partial r} - \frac{\delta(r)}{r^2}\right) = \frac{1}{l^2}\delta(r), \tag{17}$$

where $\delta(0) = 0$ by symmetry. The zero-stress conditions, $\sigma_{rr}^{(\mathrm{y})} = \sigma_{rr}^{(\mathrm{r})} = 0$ at the filament surface yield

$$\partial_r \delta(R) + \frac{\lambda}{\lambda + 2\mu}\frac{\delta(R)}{R} = 2\epsilon\frac{3\lambda + 2\mu}{\lambda + 2\mu}. \tag{18}$$

This equation is identical to the force balance equation for a finite disk in two-dimensional space studied in Ref. 38. The solution of Eqs. (17-18) can be expressed in terms of the modified Bessel functions of the first kind $I_0$ and $I_1$:

$$\delta(r) = 2\epsilon l\frac{1+\nu}{1-\nu}\frac{I_1(r/l)}{I_0(R/l) - \frac{1-2\nu}{1-\nu}\frac{I_1(R/l)}{R/l}}. \tag{19}$$

As before the yellow network displacement is the opposite of the red network displacement: $u^{(\mathrm{y})}(r) = -u^{(\mathrm{r})}(r) = \delta(r)/2$.

The surface energy of a filament of area $S_1 = 2\pi R\frac{V}{\pi R^2}$ reads

$$E_{\mathrm{surf1}} = \gamma S_1 = 2\gamma\frac{V}{R}, \tag{20}$$

leading to a total energy per unit

$$\frac{e_1(R)}{e_{\mathrm{bulk}}} = 1 - \frac{2}{3}\frac{1+\nu}{1-\nu}\left(\frac{I_1(R/l)}{(R/l)I_0(R/l) - \frac{1-2\nu}{1-\nu}I_1(R/l)} - \frac{\Gamma}{R/l}\right). \tag{21}$$

The equilibrium value of the filament radius $R_1^*$ is determined by the minimization condition $\partial e_1(R)/\partial R = 0$ at $R = R_1^*$, yielding the following equation for $s = R_1^*/l$:

$$x[I_0^2(s) - I_1^2(s)] - 2I_0(s)I_1(s) + \left[I_0(s) - \frac{1-2\nu}{1-\nu}\frac{I_1(s)}{s}\right]^2 \Gamma = 0. \tag{22}$$

Substituting the radius $R = R_1^*$ into Eq. (21) yields the minimum filament energy $e_1$.

### 3.4 Spheres

We study spherical aggregates of radius $R$ using spherical coordinates. The network displacements $u^{(\mathrm{y})}(r)$, $u^{(\mathrm{r})}(r)$ depend only on the radial coordinate $r$ and the force balance Eq. (7) reads

$$\frac{\partial^2 \delta(r)}{\partial r^2} + \frac{2}{r}\frac{\partial \delta(r)}{\partial r} - \frac{2}{r^2}\delta(r) = \frac{1}{l^2}\delta(r), \tag{23}$$

where $\delta(0) = 0$. The vanishing surface stress boundary condition Eq. (9) takes the form

$$\partial_r \delta(R) + 2\frac{\lambda}{\lambda + 2\mu}\frac{\delta(R)}{R} = 2\epsilon\frac{3\lambda + 2\mu}{\lambda + 2\mu}. \tag{24}$$

The solution to Eqs. (23-24) is

$$\delta(r) = 2\epsilon R \frac{1+\nu}{1-\nu}\frac{f(r/l)}{\sinh(R/l) - 2\frac{1-2\nu}{1-\nu}f(R/l)} \tag{25}$$

with

$$f(\tilde{r}) = \frac{1}{\tilde{r}}\left(\cosh\tilde{r} - \frac{1}{\tilde{r}}\sinh\tilde{r}\right)$$

and displacements $u^{(\mathrm{y})}(r) = -u^{(\mathrm{r})}(r) = \delta(r)/2$.

The surface energy of a sphere of area $S_0 = 4\pi R^2$ is

$$E_{\mathrm{surf0}} = \gamma S_0 = 3\gamma\frac{V}{R}. \tag{26}$$

The total energy of the sphere per unit volume $e_0(R)$ thus reads

$$\frac{e_0(R)}{e_{\mathrm{bulk}}} = 1 - \frac{1+\nu}{1-\nu}\left(\frac{f(R/l)}{\sinh(R/l) - 2\frac{1-2\nu}{1-\nu}f(R/l)} - \frac{\Gamma}{R/l}\right). \tag{27}$$

The equilibrium value of the radius $R_0^* = ls$ is given by the minimization condition $\partial e_0(R)/\partial R = 0$, or equivalently:

$$g(s) = \frac{s \cdot \sinh^2 s - s^2 f(s) \left(\frac{2}{s} \sinh s + \cosh s\right)}{\left[\sinh s - 2\frac{1-2\nu}{1-\nu} f(s)\right]^2} = \Gamma. \tag{28}$$

At a fixed Poisson's ratio, the function $g(s)$ always has a finite maximum value $g_{\max}$. For poorly deformable subunits ($\nu \geq 1/3$) $g_{\max} = 1$. For smaller $\nu$, $g_{\max}$ increases with decreasing $\nu$. As a result, the condition Eq. (28) can never be satisfied if $\Gamma > g_{\max}$. In that case, the energy per unit volume of the sphere always decreases with increasing radius and therefore has no minimum. This leads to the formation of infinite bulk aggregates or possibly other aggregate morphologies.

## 4 Morphological diagrams

To determine the most favorable aggregate morphology as a function of Poisson's ratio and surface tension, we first compute the equilibrium sizes of slabs, filaments, and spheres by solving Eqs. (14), (22) and (28) numerically. We then compare these optimal energies to each other and to the bulk energy [Eq. (6)] to determine the most stable structure for each value of the surface tension and Poisson's ratio. We also compute the energy of a spherical shell energy and optimize it with respect to its thickness and radius in the *Supplementary Information*, but find that it is never the most stable structure. We assemble these results into the morphological diagram of Fig. 3a.

We find that easily compressible subunits ($\nu < 1/3$) self-assemble into spheres at low surface tension, then transition to bulks at higher surface tension. Conversely, weakly compressible subunits ($\nu > 1/3$) successively form spheres, filaments, slabs, and finally the bulk phase as the surface tension is increased. The surface tension at the transitional line from slabs to the bulk phase is determined by the condition $\Gamma = 1$, yielding a value at the transition

$$\gamma_{2/\mathrm{b}} = \frac{1}{3}\frac{(1+\nu)}{(1-\nu)} e_{\mathrm{bulk}} l = 2\epsilon^2 \frac{\mu^{3/2}}{{\kappa_c}^{1/2}} \frac{(1+\nu)^2}{(1-\nu)^{1/2}(1-2\nu)^{3/2}}. \tag{29}$$

Very small surface tension always favors small spherical aggregates. The size of these very small spheres is moreover independent of $\nu$ (Fig. 3b). This is easily understood by noting that in the regime $R_0^* \ll l$, the individual yellow and red networks largely retain their preferred strains, while the elastic energy is almost exclusively carried by the interconnections between then. In the absence of any subunit deformation, subunit compressibility becomes irrelevant, and so does the value of $\nu$.

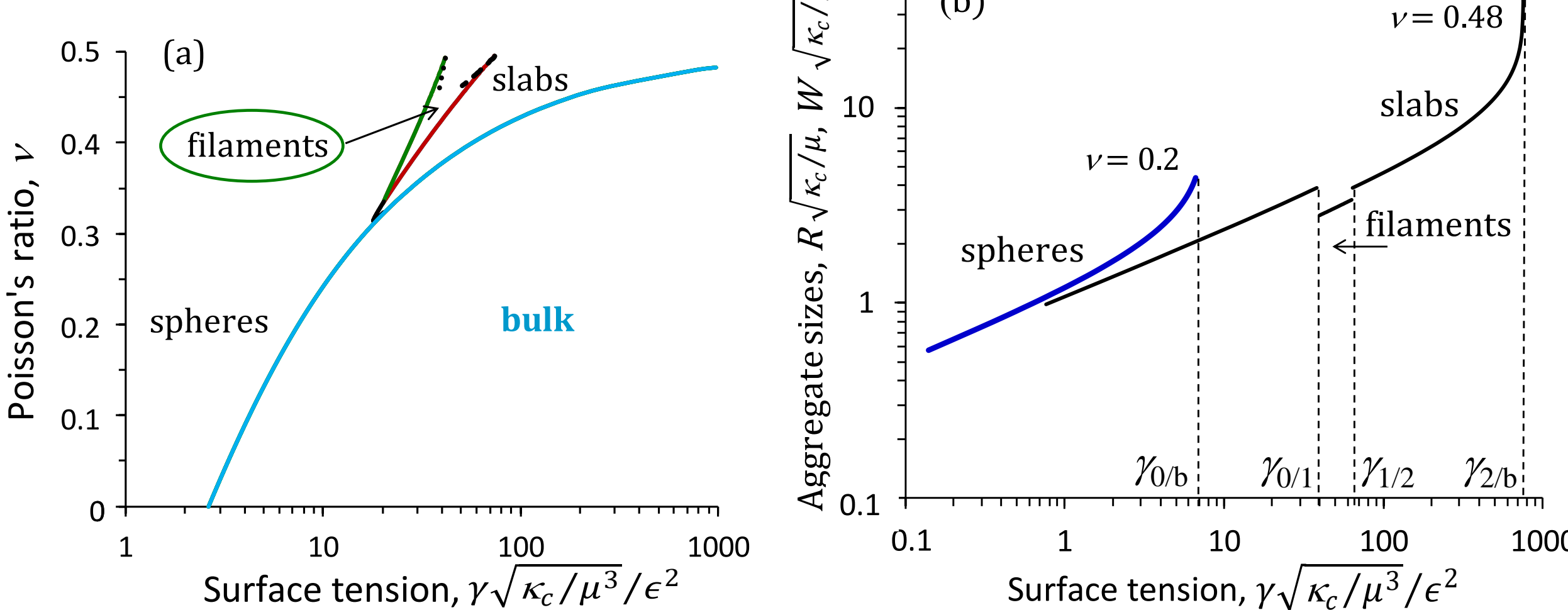

Fig. 3. (a) Morphological diagram for our aggregates. The dotted and dashed lines show the results of the asymptotic calculation of Sec. 4. (b) Aggregate sizes as a function of surfaced tension $\gamma$ at $\nu = 0.2$ (thick blue curve) and 0.48 (thin black curve). The transition values of $\gamma$ are shown by dashed lines, $\gamma_{0/\mathrm{b}} \approx 6.7\epsilon^2\sqrt{\mu^3/\kappa_c}$ at $\nu = 0.2$ and $\gamma_{2/\mathrm{b}} \approx 760\epsilon^2\sqrt{\mu^3/\kappa_c}$ ($\Gamma = 1$) at $\nu = 0.48$.

As shown in Fig. 3b, the size of the aggregates tends to increase with increasing surface tension $\gamma$. This is expected, as an increase in aggregate size is associated with a decrease in their outer surface area across the system. A closer look reveals that this increase is strictly monotonic within each type of aggregate, but that changes in aggregate dimension (*e.g.*, a transition between fiber and slab) can cause discontinuities in the aggregate size. All transitions proceed at finite aggregate size, with the exception of the transition from the slabs to the bulk phase in the small-compressibility regime $\nu > 1/3$. In this regime, the thickness of the slab diverges as it transitions to the bulk.

As $\nu$ increases, the radius of spheres decreases (Fig. 3b). This allows to decrease the effect of stress accumulation and the total energy of spheres [Eq. (27)], despite the higher surface energy. As subunits approach the incompressible limit, deforming them into a bulk becomes more and more difficult, as illustrated by the divergence of the bulk energy density $e_{\mathrm{bulk}}$ in Fig. 4. The characteristic length $l$ required for interconnecting springs to cause this deformation thus becomes ever larger. Aggregates thus increasingly find themselves in the regime of constant strain associated with aggregate sizes of the order of or smaller than $l$ (Fig. 2). We take advantage of this simplification in the asymptotic calculations of the next section.

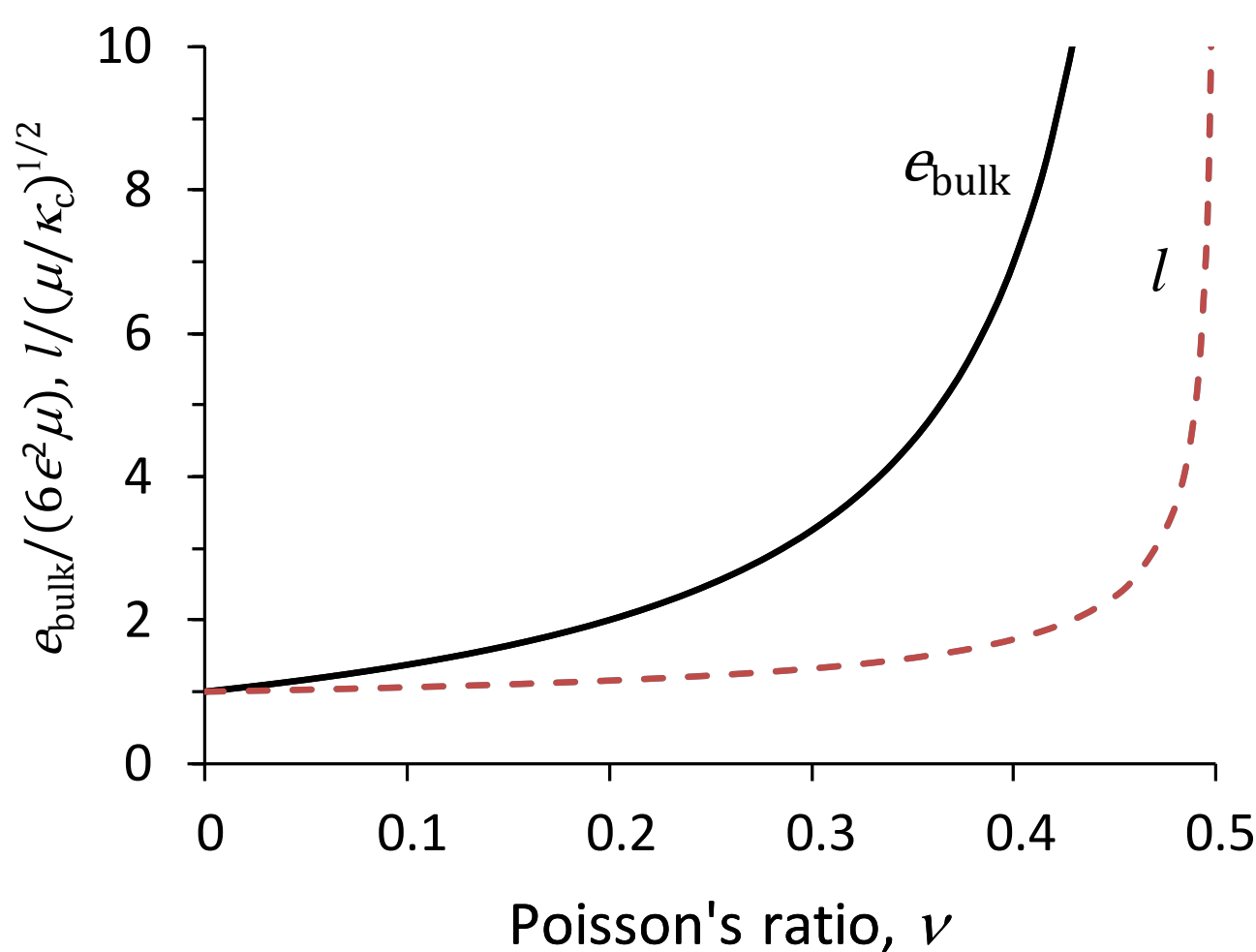

Fig. 4. Near-incompressible subunits are very difficult to deform into a bulk, as indicated by the divergence of $e_{\rm bulk}$ and $l$ as $\nu$ approaches $1/2$ at fixed $\mu$ and $\kappa_c$.

## 5 Asymptotic morphological transitions for near-incompressible subunits

To provide more transparent expressions of the implicit aggregate energies given in Sec. 3, here we derive their explicit asymptotic forms in the incompressible limit. As shown in Fig. 3(a), in this limit spheres, filaments, slabs and bulk all successively dominate as surface tension is increased. Incompressibility implies a diverging boundary layer size $l$. As a result displacement gradients $\partial_\alpha u_\beta^{(\rm y)/(\rm r)}$, which vary over a typical length scale $l$ according to Eqs. (7-8), can be approximated as constants within aggregates with dimensions much smaller than $l$.

To compute the value of these constant displacement gradients, we note that the interconnection forces of Eqs. (7-8) are negligible to leading order in $R/l$. The displacement gradients can thus be computed by minimizing elastic energy of the networks $e_{\rm y} + e_{\rm r}$ of Eqs. (2-3). The leading order of the interconnection energy can then be computed on the basis of the value of these displacement gradients.

For a flat slab, the minimum of the elastic energy $e_{\rm y} + e_{\rm r}$ [Eqs. (1-2)] corresponds to a uniform longitudinal strain $\epsilon_2 = \partial_x u_x^{(\rm y)} = -\partial_x u_x^{(\rm r)} = \epsilon \frac{3\lambda+2\mu}{\lambda+2\mu}$. The leading-order elastic energy density is $e_{\rm y2} + e_{\rm r2} = \lambda(\epsilon_2 - 3\epsilon)^2 + 2\mu[(\epsilon_2 - \epsilon)^2 + 2\epsilon^2] = 4\epsilon^2\mu\frac{3\lambda+2\mu}{\lambda+2\mu}$. The energy density of interconnection is $e_{\rm inter2}(x) = \frac{\kappa_c}{2}(2\epsilon_2 x)^2$, yielding a total interconnection energy $E_{\rm inter2} = V\frac{2}{3}\kappa_c\epsilon_2^2\left(\frac{W}{2}\right)^2$, and a surface energy $E_{\rm surf2} = \frac{\gamma V}{W/2}$. The network elastic energy does not contribute at this order of the expansion in powers of $W/l$, and neither does it in any of the asymptotic calculations in this section and the next (see *Supplementary Information*). The minimum of the total energy $E$ [Eq. (5)] respective to the slab thickness $W$ corresponds to the equilibrium value $W^* = (6\gamma/(\kappa_c\epsilon_2^2))^{1/3}$, which leads to an energy per unit volume proportional to $\gamma^{2/3}$:

$$e_2^* = 4\epsilon^2\mu\frac{3\lambda+2\mu}{\lambda+2\mu} + \left(\frac{9}{2}\kappa_c\right)^{\frac{1}{3}}(\epsilon_2\gamma)^{\frac{2}{3}}. \tag{30}$$

For a filament, the constant displacement gradient approach implies a radial and orthoradial strain $\epsilon_1 = \partial_\varphi u_\varphi^{(y)} = u_r^{(y)}/r = \partial_r u_r^{(y)}$, $-\epsilon_1 = \partial_\varphi u_\varphi^{(r)} = u_r^{(r)}/r = \partial_r u_r^{(r)}$ with $\epsilon_1 = \epsilon \frac{3\lambda+2\mu}{2(\lambda+\mu)}$. The elastic energy density of the networks reads $e_{y1} + e_{r1} = \lambda(2\epsilon_1 - 3\epsilon)^2 + 2\mu[2(\epsilon_1-\epsilon)^2 + \epsilon^2] = \epsilon^2\mu\frac{3\lambda+2\mu}{\lambda+\mu}$. The energy density of the interconnection is $e_{\text{inter1}}(x) = \frac{\kappa_c}{2}(2\epsilon_1 r)^2$, yielding a total interconnection energy $E_{\text{inter1}} = V\kappa_c\epsilon_1^2 R^2$, and a surface energy $E_{\text{surf1}} = 2\gamma V/R$. The equilibrium filament radius then reads $R_1^* = [\gamma/(\kappa_c\epsilon_1^2)]^{1/3}$ and the minimum energy of the filament aggregate per unit volume reads

$$e_1^* = \epsilon^2\mu\frac{3\lambda+2\mu}{\lambda+\mu} + 3\kappa_c^{\frac{1}{3}}(\epsilon_1\gamma)^{\frac{2}{3}}. \tag{31}$$

For a spherical aggregate, three components of the displacement gradient tensor are nonzero: $\epsilon_3 = \partial_\varphi u_\varphi^{(y)} = \partial_\theta u_\theta^{(y)} = \partial_r u_r^{(y)}$, $-\epsilon_3 = \partial_\varphi u_\varphi^{(r)} = \partial_\theta u_\theta^{(r)} = \partial_r u_r^{(r)}$. The networks can relax stress in all directions, implying that the networks relax to their preferred state $\epsilon_3 = \epsilon$ and the network energy has a minimum $e_{y0} + e_{r0} = 0$. The energy density of the interconnection is $e_{\text{inter0}}(x) = \frac{\kappa_c}{2}(2\epsilon r)^2$, which yields a total interconnection energy $E_{\text{inter0}} = V\frac{6}{5}\kappa_c\epsilon^2R^2$, and a surface energy $E_{\text{surf0}} = 3\gamma V/R$. At equilibrium the radius of the sphere is equal to $R_0^* = [5\gamma/(4\kappa_c\epsilon^2)]^{1/3}$ and its energy per unit volume reads

$$e_0^* = 9\left(\frac{1}{10}\kappa_c\right)^{\frac{1}{3}}(\epsilon\gamma)^{\frac{2}{3}}. \tag{32}$$

We note that the energy densities of all three finite aggregate morphologies are affine functions of $\gamma^{2/3}$, making it easy to compare them in Fig. 5. There we find that the most stable morphology changes successively from spheres to filaments to slabs as $\gamma$ increases, consistent with the results of Sec. 4. The surface tension values at these transitions, $\gamma_{1/2}$ and $\gamma_{0/1}$, are determined by the conditions of equal energies: $e_1^* = e_2^*$ and $e_0^* = e_1^*$ respectively. From Eqs. (30)-(32), this yields

$$\gamma_{1/2} = \left(\frac{2}{3}\right)^{3/2}\frac{(1+\nu)^2}{(1-\nu)^{3/2}}\frac{\epsilon^2}{(1-[6(1-\nu)^2]^{-1/3})^{3/2}}\sqrt{\frac{\mu^3}{\kappa_c}} \quad \text{and}$$

$$\gamma_{0/1} = \left(\frac{2}{3}\right)^{3/2}\frac{(1+\nu)^{3/2}\epsilon^2}{[3/10^{1/3} - (1+\nu)^{2/3}]^{3/2}}\sqrt{\frac{\mu^3}{\kappa_c}}. \tag{33}$$

The asymptotic transition lines are indicated by black dashed and dotted lines in Fig. 3a. The aggregates eventually transition to infinite bulk phase at even larger surface tensions, although in a regime where the approximation of constant displacement gradients breaks down as the aggregate size exceeds $l$.

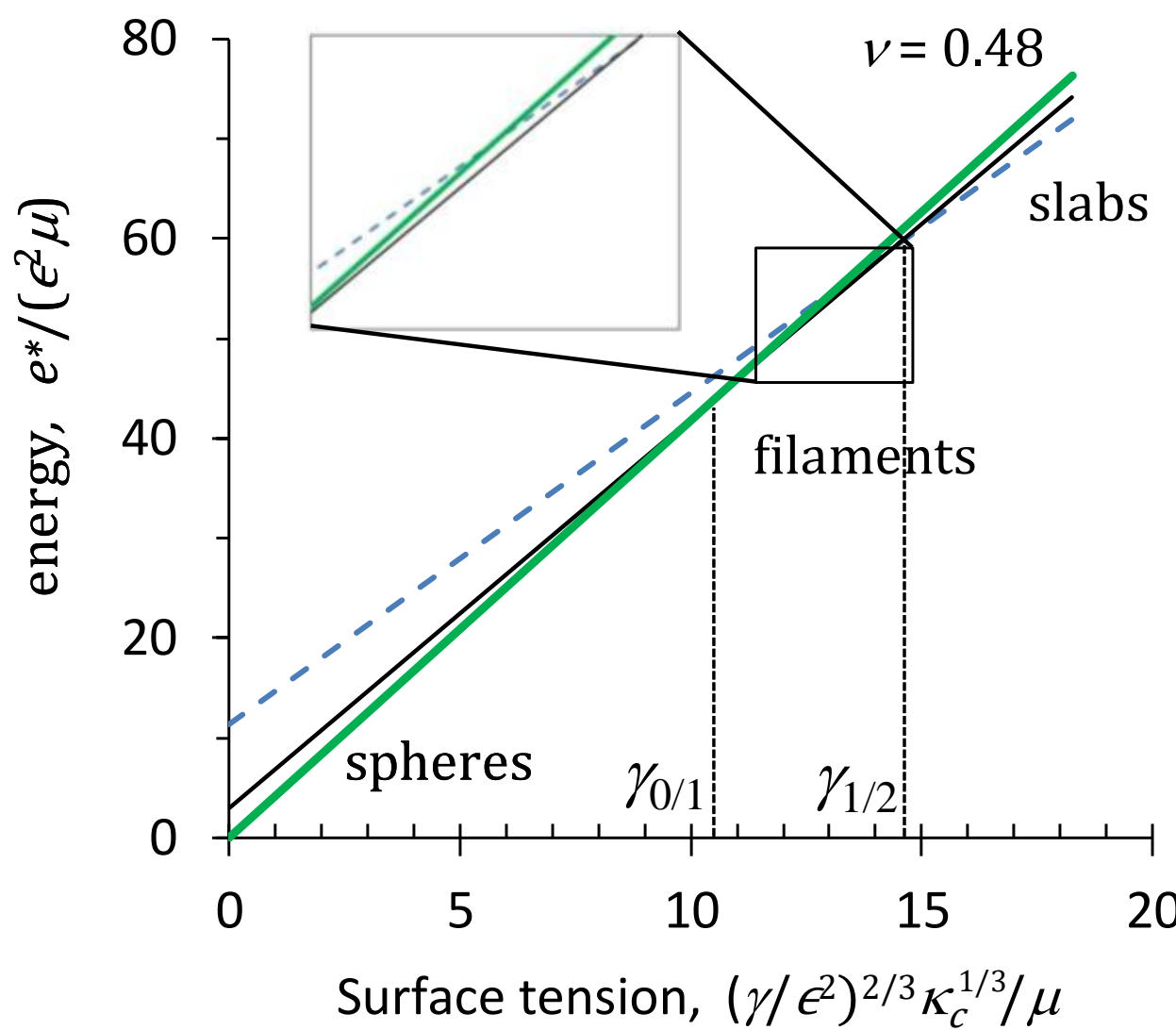

Fig. 5. Equilibrium energies per unit volume $e^*$ for spherical aggregates (thick line), filaments (thin solid line), and slabs (dashed line) within the constant displacement gradients approximation valid for almost-incompressible subunits $\nu \simeq 1/2$. Here we plot these lines for $\nu = 0.48$. For each value of the surface tension, the most energetically favorable aggregate morphology corresponds to the lowest line. The filament region is between the transition values $\gamma_{0/1}$ and $\gamma_{1/2}$. The inset outlines the region where filaments dominate.

## 6 Asymptotic behavior of aggregates in $D$-dimensional space

To determine whether the successive transitions between aggregates of increasing dimensionality are a generic feature of low-compressibility subunits in our model, we investigate the near-incompressible limit of our model in arbitrary spatial dimension $D$. The elastic energy densities of the yellow and red networks generalize to

$$e_{\mathrm{y}} = \frac{\lambda}{2}\left(\partial_\alpha u_\alpha^{(\mathrm{y})} - D\epsilon\right)^2 + \mu\left(\frac{\partial_\alpha u_\beta^{(\mathrm{y})} + \partial_\beta u_\alpha^{(\mathrm{y})}}{2} - \epsilon\delta_{\alpha\beta}\right)^2, \tag{34}$$

$$e_{\mathrm{r}} = \frac{\lambda}{2}\left(\partial_\alpha u_\alpha^{(\mathrm{r})} + D\epsilon\right)^2 + \mu\left(\frac{\partial_\alpha u_\beta^{(\mathrm{r})} + \partial_\beta u_\alpha^{(\mathrm{r})}}{2} + \epsilon\delta_{\alpha\beta}\right)^2, \tag{35}$$

where the spatial coordinates $\alpha, \beta$ now run over the coordinates $\{x_1, x_2, \ldots, x_D\}$ of $D$-dimensional space. The form of Eq. (3), which describes the interconnection energy, is unchanged. The $D$-dimensional Poisson's ratio $\nu = \lambda/[\lambda(D-1) + 2\mu]$ must be smaller or equal to $1/(D-1)$, which corresponds to the incompressible limit. As previously we use the $D$-dimensional bulk as our reference state. Its energy per unit volume reads

$$e_{\mathrm{bulk}} = \epsilon^2 D\lambda\left(D + \frac{2\mu}{\lambda}\right) = 2\epsilon^2 D\mu\frac{1+\nu}{1-(D-1)\nu}. \tag{36}$$

We consider symmetric aggregates unlimited in $d$ spatial directions. Their size is limited to a radius $R_d$ in the remaining $D-d$ directions, so that $\sum_{\alpha=d+1}^{D-d} x_\alpha^2 \in [0, R_d^2]$. As an illustration, in the previously studied $D = 3$ case $R_2$ is the half-width of the slab, $R_1$ is the filament radius, and $R_0$ is the sphere radius. As previously, the longitudinal displacement gradients along the unlimited directions lest the energy of the network interconnection diverge: $\forall\alpha \in [1, d]\;\; \partial_\alpha u_\alpha^{(\mathrm{y})/(\mathrm{r})} = 0$. The non-diagonal elements of the displacement gradient vanish by symmetry: $\partial_\alpha u_{\beta\neq\alpha}^{(\mathrm{y})/(\mathrm{r})} = 0$.

As in Sec. 4 we consider the asymptotic regime $R_d \ll l$. This regime dominates almost-incompressible systems, for which $l$ is very large. Note that it also applies for small surface tension $\gamma$ irrespectively of the value of Poisson's ratio. In this approximation, the energy of the network interconnection is small and, as a result, the non-zero strain components are approximately constant and equal to each other. We denote the value of this strain by $\epsilon_d$, implying that for $\alpha > d$ we have $\partial_\alpha u_\beta^{(\mathrm{y})} = -\partial_\alpha u_\beta^{(\mathrm{r})} = \epsilon_d \delta_{\alpha\beta}$. The elastic energies of the red and yellow networks are identical and their sum reads

$$e_{\mathrm{y}\,d} + e_{\mathrm{r}\,d} = \lambda[(D-d)\epsilon_d - D\epsilon]^2 + 2\mu[(D-d)(\epsilon_d - \epsilon)^2 + d\epsilon^2]\,. \tag{37}$$

The value of the equilibrium strain $\epsilon_d$ is obtained by minimizing the elastic energy with respect to it, yielding

$$\epsilon_d = \epsilon\frac{D + 2\mu/\lambda}{D - d + 2\mu/\lambda} = \epsilon\frac{1+\nu}{1-(d-1)\nu}. \tag{38}$$

Substituting this result into Eq. (37) we find

$$e_{\mathrm{y}\,d} + e_{\mathrm{r}\,d} = 2\epsilon^2\mu d\frac{D + 2\mu/\lambda}{D - d + 2\mu/\lambda}. \tag{39}$$

The displacement fields in our symmetric aggregates depend linearly on the radial coordinate of the $(D-d)$-dimensional sphere, namely $u_r^{(\mathrm{y})} = -u_r^{(\mathrm{r})} = \epsilon_d r$. The energy of the network interconnection thus reads

$$e_{\mathrm{inter}\,d} = \frac{1}{V_{D-d}R_d^{D-d}}\frac{\kappa_c}{2}\int_0^{R_d} dr\, S_{D-d}r^{D-d-1}(2\epsilon_d r)^2, \tag{40}$$

where $V_{D-d}$ is the volume of a sphere of unit radius in dimension $D-d$ and $S_{D-d} = (D-d)V_{D-d}$ its surface area. Using Eq. (38) we obtain

$$e_{\mathrm{inter}\,d} = 2\kappa_c\epsilon^2\frac{D-d}{D-d+2}\left[\frac{1+\nu}{1-(d-1)\nu}\right]^2 R_d^2. \tag{41}$$

To compute the optimal aggregate radius, we introduce the surface energy $E_{\mathrm{surf}} = \gamma S$, where $S$ is the aggregate surface area. As in the previous sections, the surface-to-volume ratio of our aggregates is equal to the surface-to-volume ratio of the sphere in dimension $D-d$, namely $(D-d)/R_d$. The surface energy per unit volume thus reads

$$e_{\mathrm{surf}\,d} = \gamma\frac{D-d}{R_d}\,. \tag{42}$$

Adding the network, interconnection and surface energies of Eqs. (39), (41-42) we obtain the total energy

$$\frac{e_d(R_d)}{e_{\text{bulk}}} = 1 - C_d\left[1 - \frac{1}{D-d+2}\frac{1-(D-2)\nu}{1-(d-1)\nu}\left(\frac{R_d}{l}\right)^2 - \frac{1-(d-1)\nu}{1-(D-2)\nu}\frac{\Gamma}{R_d/l}\right], \quad (43)$$

where

$$C_d = \frac{(D-d)(1+\nu)}{D[1-(d-1)\nu]} \quad (44)$$

and

$$\Gamma = \frac{D[1-(D-2)\nu]}{1+\nu}\frac{\gamma}{l e_{\text{bulk}}} \quad (45)$$

is the $D$-dimensional generalization of the dimensionless surface tension $\Gamma$ defined in Eq. (15).

In the case of an aggregate limited in one direction only ($d = D-1$), *i.e.* a slab in $D$-dimensional space, the aggregate energy for any value of the slab thickness can be derived similarly to three-dimensional case (Sec. 3.1). Equation (14) is thus a special case of Eq. (43) with radius $R_{D-1} = W/2$. Therefore, the transition from slabs in $D$-dimensional space to the bulk phase occurs for $\Gamma = 1$, as for $D = 3$. The bulk energy $e_{\text{bulk}}$ and the characteristic length $l$ diverge in the incompressible limit $\nu \to 1/(D-1)$, and as a result the surface tension $\gamma_{(D-1)/\text{b}}$ associated with this transition also diverges:

$$\gamma_{(D-1)/\text{b}} = \frac{1}{D}\frac{(1+\nu)}{[1-(D-2)\nu]}e_{\text{bulk}}l = 2\epsilon^2\frac{\mu^{3/2}}{{\kappa_c}^{1/2}}\frac{(1+\nu)^2}{[1-(D-2)\nu]^{1/2}[1-(D-1)\nu]^{3/2}}. \quad (46)$$

For spherical aggregates, *i.e.* the $d = 0$ case, stresses can be relieved in all $D$ dimensions. Equation (38) give a constant strain $\epsilon_d = \epsilon$ and resulting in a vanishing network energy $e_{\text{y}\,d} + e_{\text{r}\,d} = 0$ in Eq. (39). The network energies for the other aggregate morphologies are strictly positive, implying that spheres are the most favorable morphologies at small $\gamma$ regardless of Poisson's ratio.

We now tackle the case of general $d$-dimensional aggregates. We minimize the total volumic energy $e_d$ of Eq. (43) with respect to the aggregate radius $R_d$ and obtain the equilibrium aggregate size

$$\frac{R_d^*}{l} = \left(\frac{D-d+2}{2}\Gamma\right)^{1/3}\left[\frac{1-\nu(d-1)}{1-\nu(D-2)}\right]^{2/3} \quad (47)$$

and the energy per unit volume

$$\frac{e_d^*}{e_{\text{bulk}}} = (1 - C_d) + C'_d\Gamma^{2/3}, \quad (48)$$

where

$$C'_d = 3(1+\nu)\frac{D-d}{D}\{4(D-d+2)[1-\nu(D-2)][1-\nu(d-1)]^2\}^{-1/3}. \quad (49)$$

Regardless of the aggregate dimensionality $d$, the dependence of the energy $e_d^*$ of Eq. (48) on the surface tension $\Gamma$ is characterized by the same exponent 2/3 as in three-dimensional space. As a result, the relative ordering of the aggregate energies depends solely on the values of the coefficients $C_d$ and $C'_d$. Consistent with our previous argument, we find that the $D$-dimensional spherical aggregate $d=0$ is the most stable at smaller surface tension, and that the system undergoes a sequence of transitions whereby $d$ increases by one unit for values of the dimensionless surface tension given by

$$\Gamma_{d/(d+1)} = \left(\frac{C_d - C_{d+1}}{C'_d - C'_{d+1}}\right)^{3/2}. \quad (50)$$

The corresponding transition values for the surface tension $\gamma_{d/(d+1)}$ can be then found using Eq. (45). We illustrate these transitions for four-dimensional space in Fig. 6a and in arbitrary spatial dimension $D$ in Fig. 6b.

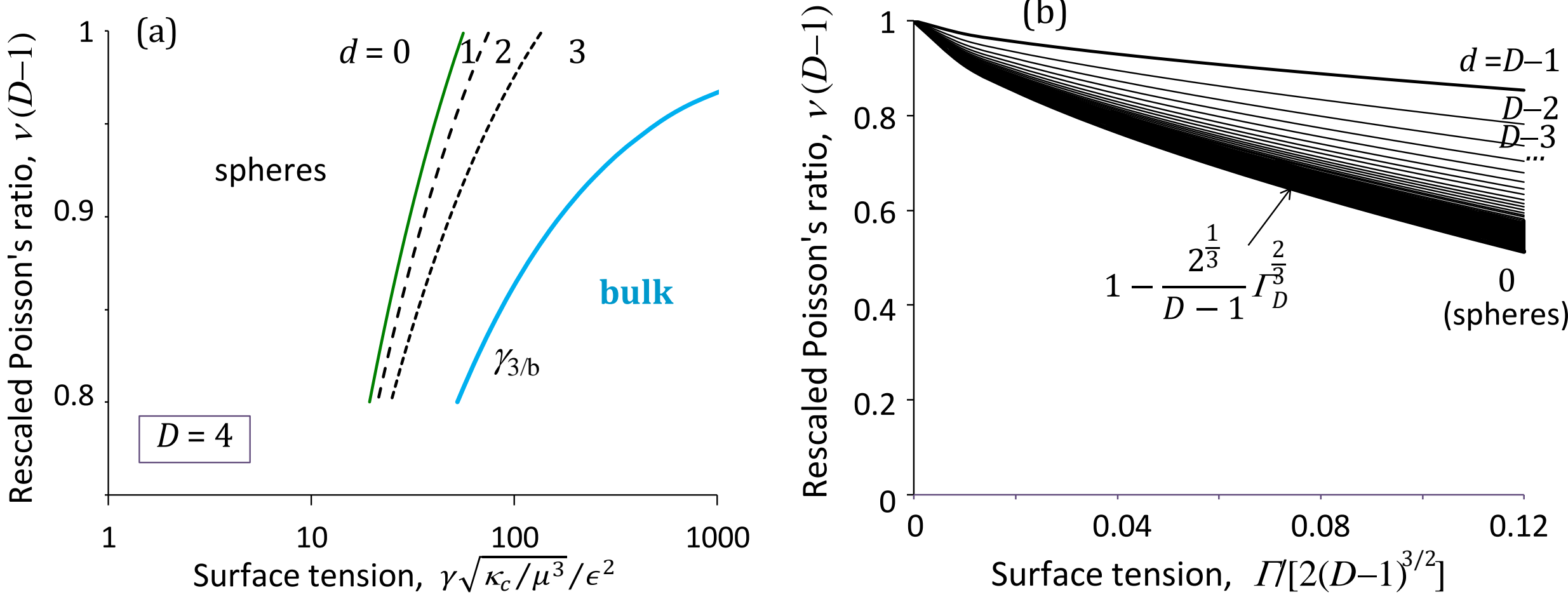

Fig. 6. Asymptotic transition curves between domains of aggregates unlimited in increasing number of dimensions (a) in four-dimensional space in the coordinates $\nu$ and $\gamma$, (b) in the general $D$-dimensional case in the coordinates $\nu$ and $\Gamma$. Here we rescale Poisson's ratio $\nu$ by its value at incompressibility $1/(D-1)$ and the dimensionless surface tension $\Gamma$ by $2(D-1)^{3/2}$ as indicated on the axes. The surface tension $\gamma_{3/\mathrm{b}}$ at the transition from slabs to the bulk phase for $D=4$ is determined by Eq. (46). In the coordinates of Fig. 6b, the position of the transition curve between $D-i$ and $(D-i-1)$-dimensional aggregates depends only on $i$ but not on $D$. As a result, the same diagram is valid regardless of the value of $D$ up to the removal of all curves with $i \geq D$. The asymptotic position ($D \to \infty$) of the last transition to $D$-dimensional spheres is shown by the arrow.

## 7 Discussion

Ill-fitting particles have a propensity to aggregate into symmetry-broken, partially self-limited aggregate. However the mechanisms that lead to the choice of one aggregate dimensionality over another are still not fully understood. Here we provide the complete morphological diagram for this behavior in three dimensions, as well as insights into higher-dimensional behaviors. In our model aggregate self-limitation is controlled by an emergent length scale built from a combination of elastic properties of the subunits.

We find that symmetry breaking upon aggregating occurs in particles with low compressibility, but not in very compressible ones. This can be understood by recalling that Poisson's ratio characterizes the ability of a material to redirect stresses exerted in one direction into perpendicular directions. At sufficiently high $\nu$, strains in the directions along which the aggregate is infinite also generates stresses in other directions, which allows these stresses to be partially relieved. Thus, we observe the self-assembly of high-symmetry subunits into aggregates of all possible dimensions. This behavior is reminiscent of our previous two-dimensional findings [38], and here we additionally establish that self-assembly in higher dimension can proceed through a sequence of aggregates of increasing dimension as the aggregate surface tension is increased.

Our work considers highly symmetric subunits both for simplicity and to highlight the possibility of spontaneous symmetry breaking upon aggregation. Asymmetric subunits are likely to facilitate the formation of lower-symmetry aggregates [27,28,31,32]. They however require more complicated descriptions than the minimal model proposed here. Our framework could be extended to describe such cases, for instance by introducing more than two deformation fields, which will result in a larger number of couplings. Such cases are likely to involve multiple elastic boundary layer lengths instead of the single one studied here, thus possibly giving rise to a more complicated phenomenology. While our formalism concentrates on simple aggregate structures in the absence of thermal noise, our previous two-dimensional Monte-Carlo simulations suggest that our approach is robust to the introduction of more complicated, fluctuating morphologies. Finally, although our continuum approach formally requires that the boundary layer length be much larger than the subunit size, these simulations demonstrate that its predictions can remain valid even when they are of the same order [38].

Our insights into the generic behavior of aggregating ill-fitting objects could be applied to specific, more realistic models of protein elasticity [39,40] to understand the drivers of their aggregation into fibers during disease. Parameters such as the surface tension $\gamma$ should be easily accessible experimentally, *e.g.*, by introducing depletion interactions to drive protein aggregation, and targeted mutations or protein design could be harnessed to selectively alter the proteins' elasticity [41]. In addition, the same self-assembly principles are likely to apply in artificial systems, *e.g.*, based on colloids or DNA origami [13,17,22,36]. The principles outlined here could thus be used as an economic design for self-limiting assembly on large length scales.

## Conflicts of interest

There are no conflicts to declare.

## Data availability

Neither experimental nor simulation data were produced during this work. An additional description of the model of spherical shell aggregates, an alternative representation of the morphological diagram, and a discussion of the calculation accuracy for asymptotic energies have been included as Supplementary Information.

## Acknowledgements

This work was supported by ANR Grants No. ANR-21-CE11-0004-02, No. ANR-22-ERCC-0004-01, and No. ANR-22-CE30-0024-01, as well as ERC Starting Grant No. 677532 and the Impulscience® program from Fondation Bettencourt-Schueller. M.L.'s group belongs to the CNRS consortium AQV.

# Supplementary Information

## 1. Aggregates in the form of spherical shell

We consider an aggregate in the shape of a spherical shell, or a vesicle, with the outer radius $R$ and the inner radius $R-\Delta R$. We assume a spherical symmetry of the displacements $u^{(\mathrm{y})}(r)$ and $u^{(\mathrm{r})}(r)$ of the red and yellow networks and use the spherical coordinates with the center in the centre of the vesicle. The force balance equation for the difference $\delta(r)=u^{(\mathrm{y})}(r)-u^{(\mathrm{r})}(r)$ in the spherical coordinates is the same as for a spherical aggregate (Eq. (18)):

$$\frac{\partial^2\delta(r)}{\partial r^2}+\frac{2}{r}\frac{\partial\delta(r)}{\partial r}-\frac{2}{r^2}\delta(r)=\frac{1}{l^2}\delta(r). \qquad (S1)$$

The conditions of the zero stress at the inner and outer surfaces, $\sigma_{rr}^{(\mathrm{y})}=\sigma_{rr}^{(\mathrm{r})}=0$ at $r=R-\Delta R$ and $R$, yield

$$\partial_r\delta(R)+2\frac{\lambda}{\lambda+2\mu}\frac{\delta(R)}{R}=2\epsilon\frac{3\lambda+2\mu}{\lambda+2\mu} \qquad (S2)$$

$$\partial_r\delta(R-\Delta R)+2\frac{\lambda}{\lambda+2\mu}\frac{\delta(R-\Delta R)}{R-\Delta R}=2\epsilon\frac{3\lambda+2\mu}{\lambda+2\mu} \qquad (S3)$$

The solution for eqs. (S1)-(S3) is

$$\delta(r)=2\epsilon R\frac{1+\nu}{1-\nu}\frac{f_{\mathrm{shell}}(r/l)}{\left(1+4\frac{\mu}{M}\frac{l^2}{R^2}\right)(U_2-U_1)-4\frac{\mu}{M}\frac{l}{R}(U_2+U_1)}, \qquad (S4)$$

$$f_{\mathrm{shell}}(x)=U_2\left(\frac{1}{x}-\frac{1}{x^2}\right)e^{-\left(\frac{R}{l}-x\right)}+U_1\left(\frac{1}{x}+\frac{1}{x^2}\right)e^{\frac{R}{l}-x},$$

where $M=2\mu\frac{1-\nu}{1-2\nu}$ is the P-wave or longitudinal modulus, $\theta=\Delta R/R$,

$$U_1=\frac{e^{-\theta R/l}}{1-\theta}-1-4\frac{\mu}{M}\frac{l}{R}\left(\frac{e^{-\theta R/l}}{(1-\theta)^2}-1-\frac{l}{R}\left(\frac{e^{-\theta R/l}}{(1-\theta)^3}-1\right)\right), \qquad (S5)$$

$$U_2=\frac{e^{\theta R/l}}{1-\theta}-1+4\frac{\mu}{M}\frac{l}{R}\left(\frac{e^{\theta R/l}}{(1-\theta)^2}-1+\frac{l}{R}\left(\frac{e^{\theta R/l}}{(1-\theta)^3}-1\right)\right).$$

The displacements of the red and yellow networks $u^{(\mathrm{y})}(r)=-u^{(\mathrm{r})}(r)=\delta(r)/2$.

The surface energy of the spherical shell of area $S_{\mathrm{sh}}=4\pi(R^2+(R-\Delta R)^2)=\alpha_S 4\pi R^2$ and volume $V_{\mathrm{shell}}=\frac{4}{3}\pi(R^3-(R-\Delta R)^3)=\alpha_{\mathrm{shell}}\frac{4}{3}\pi R^3$ is

$$E_{\mathrm{surf\,sh}}=\gamma S_{\mathrm{sh}}=3\frac{\alpha_S}{\alpha_{\mathrm{shell}}}V_{\mathrm{shell}}\frac{\gamma}{R} \qquad (S6)$$

From Eq. (1) for the elastic energy and eqs. ($S4$)-($S6$), we find the energy of the spherical shell per unit volume $e_{\text{shell}}(R)$:

$$\frac{e_{\text{shell}}}{e_{\text{bulk}}} = 1 - \frac{1+\nu_3}{1-\nu_3}\frac{1}{\alpha_{\text{shell}}}\left(\frac{f_{\text{shell}}(R/l) - (1-\theta)^2 f_{\text{shell}}((1-\theta)R/l)}{(U_2 - U_1)\left(1 + \frac{1}{\tilde{R}^2}4\frac{\mu}{M}\right) - \frac{1}{\tilde{R}}4\frac{\mu}{M}(U_2 + U_1)} - \alpha_S\frac{\Gamma}{\tilde{R}}\right) \qquad (S7)$$

Numerical analysis of the shell energy ($S7$) at given values of the elastic moduli and dimensionless surface tension $\Gamma$ predicts that the formation of a solid spherical aggregate is energetically more favorable than the formation of a shell.

## 2. Aggregate morphology depending on normalized surface tension $\Gamma$

A morphological diagram of the aggregates can be presented in the coordinates $\Gamma$ and $\nu$, where $\Gamma = 3\gamma(1-\nu)/[e_{\text{bulk}}l(1+\nu)]$ is the normalized surface tension (Fig. S1). This diagram describes the same transitions as the diagram in the coordinates $\Gamma$ and $\nu$ in Sec. 4 (Fig. 3a), however its general view is different. Since the normalized surface tension $\Gamma \to 0$ far a finite value of $\gamma$ in the incompressible limit $\nu \to 1/2$, the transition curves between spheres and filaments, as well as between filaments and slabs on the diagram (Fig. S1) start from the same point $\Gamma = 0$, $\nu = 1/2$. The vertical transition line between slabs and infinite bulk, $\Gamma = 1$, corresponds to equal energies of the surface and bulk for a slab of half thickness $l$.

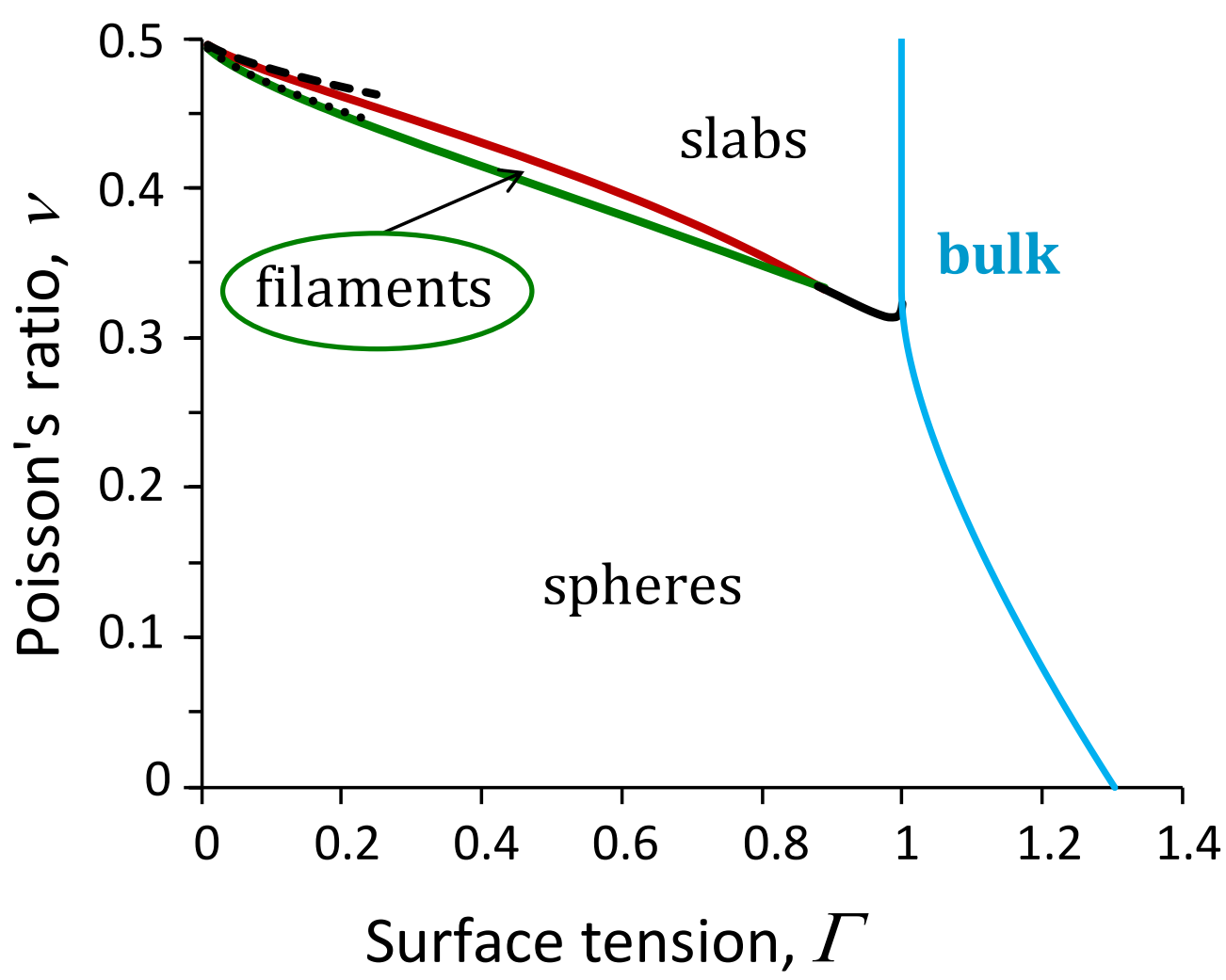

Fig. S1. Morphological diagram of the aggregates of ill-fitting subunits in 3D-space in the coordinates $\Gamma$ and $\nu$, where $\Gamma = 3\gamma(1-\nu)/[e_{\text{bulk}}l(1+\nu)]$. The asymptotic transition lines spheres/filaments and filaments/slabs valid in the incompressible limit $\nu \to 1/2$ (Sec. 5) are shown by the dotted and dashed lines, respectively.

By definition the dimensionless surface tension $\Gamma$ characterizes the ratio between the surface energy of a slab of half-thickness $l$ and its bulk elastic energy in the reference state [Eq. (15)]. Phase separation to form the bulk phase becomes possible when the surface energy exceeds the bulk energy, $\Gamma > 1$. At $\nu < \nu_{cr}$, the value of Poisson's ratio at the transition spheres-bulk decreases (Fig. 3) with increasing the surface tension $\Gamma$ and increases with increasing $\tilde{\gamma}$.

A morphological diagram in similar coordinates was presented for 2D aggregates in Ref. 38. The morphologies considered there are discs, strips, and the two-dimensional bulk phase.

## 3. Asymptotic energy for aggregates of near-incompressible subunits

In Sec 5 we analyze asymptotic morphological transitions between slabs and filaments and between filaments and spheres in the limit of highly incompressible subunits ($\nu \to 1/2$). In this case the characteristic length scale $l$ diverges (Fig. 4) and displacement gradients $\partial_\alpha u_\beta^{(\mathrm{y})/(\mathrm{r})}$ can be approximated as constants (from Eqs. (7) and (8)), as the aggregate size $R$ is much smaller than $l$. This yields a constant value as a dominating term in the energy density of the elastic networks, that can be written in the form $e_{\mathrm{y}d} + e_{\mathrm{r}d} = \alpha_d \epsilon^2 \mu$, where $\alpha_d$ is a numerical coefficient of the order of unity [in Eqs. (30-32)] for an aggregate unlimited in $d = 1$ or 2 directions and $\alpha_0 = 0$.

Deviation of the displacement gradients from the approximating constant values are of the order of $\frac{R}{l}$, yielding [Eq. (1),(2) and (5)] the correction of the same order to the network energy: $\frac{1}{V}\int dV\left(\Delta e_{\mathrm{y}d} + \Delta e_{\mathrm{r}d}\right) = \alpha'_d \epsilon^2 \mu \frac{R}{l}$, where $\alpha'_d$ is a numerical coefficient of the order of unity at any values of $\mu$ and $\nu$. Below we explain why this correction can be neglected relatively to other energy contributions.

Under the condition of constant displacement gradients, the interconnection energy can be represented as

$$\frac{E_{\mathrm{interd}}}{V} = \beta_d \kappa_c \epsilon^2 R^2, \qquad (S8)$$

where $\beta_d$ is a numerical coefficient of the order of unity. Using the definition (11) of the characteristic scale $l$, we can rewrite the expression ($S$8) as

$$\frac{E_{\mathrm{interd}}}{V} = \beta_d \mu \epsilon^2 \frac{1-\nu}{1-2\nu}\left(\frac{R}{l}\right)^2. \qquad (S9)$$

Despite this contribution is proportional to a small parameter $\left(\frac{R}{l}\right)^2$, it contain $1-2\nu$ in the denominator. In the limit of low compressibility, $(1-2\nu) \to 0$ that makes the interconnection energy comparable to the main term of the network energy for $d = 1,2$. For spheres ($d = 0$), $e_{\mathrm{y}d} + e_{\mathrm{r}d} = 0$ in the main approximation and then interconnection energy dominates over the next order terms.

The surface energy per unit volume

$$\frac{E_{\mathrm{surf}d}}{V} = \frac{(D-d)\gamma}{R} \qquad (S10)$$

is inversely proportional to the aggregate size irrespectively of the strain distribution in the aggregates.

Therefore, the corrections to the network energy $e_{\mathrm{y}d} + e_{\mathrm{r}d}$ originated from the strain deviations from linearity, would not change the transition values of the surface tension $\gamma$ found in Sec. 5 in the main approximation.